\documentclass[epj]{webofc}
\usepackage[varg]{txfonts}   

\newcommand{\order}{{\cal O}}
\newcommand{\One}{1\kern-4.5pt1}

\newcommand{\mrm}[1]{\mathrm{#1}}

\newcommand{\om}{\omega}

\newcommand{\ncfg}{N_{\text{cfg}}}

\wocname{EPJ Web of Conferences}
\woctitle{CONF12}
\begin{document}
\selectlanguage{english}
\title{Thermal D mesons from anisotropic lattice QCD}
%
%

\author{Aoife Kelly\inst{1} \and
Jon-Ivar Skullerud\inst{1}\fnsep\thanks{\email{jonivar@thphys.nuim.ie}}
}

\institute{Department of Mathematical Physics, National University of
  Ireland Maynooth, Maynooth, Co Kildare, Ireland}


\abstract{%
We present results for correlators and spectral functions of
  open charm mesons using 2+1 flavours of clover fermions on
  anisotropic lattices. The D mesons are found to dissociate close to the
  deconfinement crossover temperature $T_c$. Our preliminary results
  suggest a shift in the thermal D meson mass below $T_c$.  Mesons
  containing strange quarks exhibit smaller thermal modifications than
those containing light quarks.
}
\maketitle
\section{Introduction}
\label{intro}

The study of heavy quarks, and in particular charm quarks, in high
temperature QCD has a long history.  Until recently, however, the
focus has been almost exclusively on quarkonia, ie the $c\bar{c}$ and
$b\bar{b}$ systems.  This has changed in the past few years, with an
increased experimental and theoretical interest in open charm,
including D meson flow \cite{Abelev:2013lca} and yields
\cite{Adamczyk:2014uip,Adam:2015sza,Adam:2015jda}.

There are several reasons why it is important to study open charm
alongside charmonium.  Since charm quarks are not created or destroyed
thermally to any significant degree at the temperatures reached at
RHIC and LHC, any decreased yield of charmonium states must be
associated with an increased yield in open charm (although this can be
hard to identify experimentally).  It has also been suggested that the
double ratio $R_{AA}(J/\psi)/R_{AA}(D)$ may be a better measure of
medium modifications than the traditional $R_{AA}(J/\psi)$ as a number
of systematic effects including cold nuclear matter effects cancel out
in this ratio.

An important issue to understand in this context is to what extent
open charm mesons experience thermal modifications below $T_c$, and
whether some bound states may survive above $T_c$.  It has for example
been suggested that the abundance of $D_s$ states may increase
relative to the states containing light quarks.

There have been very few lattice studies of open charm at high
temperature so far
\cite{Bazavov:2014yba,Bazavov:2014cta,Mukherjee:2015mxc}.  These have
used 
spatial correlators and cumulants to assess the possible survival of
open charm bound states above $T_c$.
Here we will for the first time present results for temporal
correlators and spectral functions of open charm mesons.

\section{Methods}
\label{methods}

This study uses the FASTSUM collaboration
ensemble~\cite{Aarts:2014cda,Aarts:2014nba}, with 2+1 flavours of
anisotropic clover fermions and a mean-field improved anisotropic
Symanzik gauge action.  The spatial lattice spacing is $a_s=0.123$~fm
and the anisotropy $\xi=a_s/a_\tau=3.5$.  The strange quark mass is
tuned to its physical value, while the light quarks correspond to a
pion mass $m_\pi\approx380$~MeV.  Further details about the ensembles
are given in table~\ref{tab:lattices}.  The action is identical to
that used by the Hadron Spectrum Collaboration~\cite{Edwards:2008ja},
and the zero temperature ($N_\tau=128$) configurations were kindly
provided by them.  For the charm quarks, we have also used the
parameters from the Hadron Spectrum Collaboration~\cite{Liu:2012ze}.
Both the configurations and the correlators were generated using the
Chroma software package~\cite{Edwards:2004sx}.
\begin{table}
\centering
\caption{Lattice volumes $N_s^3\times N_\tau$, temperatures $T$ and
  number of configurations $\ncfg$ used in this work.  The
  pseudocritical temperature $T_c$ was determined from the inflection
  point of the Polyakov loop~\cite{Aarts:2014nba}.}
\label{tab:lattices}
\begin{tabular}{|ccccr|}\hline
$N_s$&   $N_\tau$ & $T$ (MeV) & $T/T_c$ & $\ncfg$ \\\hline
24 & 128 & \,44 & 0.24 &  500 \\
24 &  40 & 141  & 0.76 &  500 \\
24 &  36 & 156  & 0.84 &  500\\
24 &  32 & 176  & 0.95 & 1000 \\
24 &  28 & 201  & 1.09 & 1000 \\
24 &  24 & 235  & 1.27 & 1000 \\
24 &  20 & 281  & 1.52 &  576 \\
24 &  16 & 352  & 1.90 & 1000 \\ \hline
\end{tabular}
\end{table}

Information about hadronic states (including energies, widths and
thresholds) in the medium is contained in the spectral function
$\rho(\om;T)$, which for mesonic states is related to the euclidean
correlator $G(\tau;T)$ according to
\begin{equation}
G(\tau;T) =
\int\rho(\omega;T)K(\tau,\omega;T)d\omega\,, \quad
K(\tau,\omega;T) = \frac{\cosh[\omega(\tau-1/2T)]}{\sinh(\omega/2T)}\,,
\label{eq:spectral}
\end{equation}
Determining
$\rho(\om)$ from a given (noisy) $G(\tau)$ cannot be done exactly; we
will here employ the maximum entropy method (MEM) to obtain the most
likely spectral function for the given data.  The spectral function is
written in terms of a default model $m(\om)$, which encodes prior
information, and a set of $N_b$ basis functions $u_k(\om)$ as
\begin{equation}
 \rho(\om) = m(\om)\exp[\sum_{k=1}^{N_b}b_ku_k(\om)]\,.
\label{eq:spectral-expansion}
\end{equation}
The standard implementation of MEM employs the singular value
decomposition (SVD) of the kernel $K$,
\begin{equation}
 K(\om_i,\tau_j) = K_{ij} = (U\Xi V)_{ji}\,,
\label{eq:svd}
\end{equation}
and the basis functions $u_k$ are chosen to be the column vectors of
$U$ corresponding to the $N_s\leq N_{\text{data}}$ singular values in
the diagonal matrix $\Xi$.  We will instead use a Fourier basis
\cite{Rothkopf:2012vv} which was found to yield more reliable results
for the data at hand.  The results have been cross-checked using the
standard SVD basis and the SVD basis with extended search
space~\cite{Rothkopf:2011ef}, and found to be consistent within the
uncertainties inherent in the respective methods.

The systematic uncertainty of the MEM can be avoided by studying the
reconstructed correlator, defined as
\begin{equation}
 G_r(\tau;T,T_r) = \int_0^\infty\rho(\om;T_r)K(\tau,\om,T)d\om\,,
\end{equation}
where $T_r$ is a reference temperature where the spectral function can
be reliably constructed, usually chosen to be the lowest available
temperature.  It is clear that if there are no medium modifications,
ie $\rho(\om;T)=\rho(\om;T_r)$ then $G_r(\tau;T,T_r)=G(\tau;T)$
(although the converse is not necessarily the case).

The reconstructed correlator can also be computed directly from the
underlying correlator $G(\tau;T_r)$ without having to extract any
spectral functions~\cite{Ding:2012sp}.  Using
\begin{equation}
\frac{\cosh\big[\om(\tau-N/2)\big]}{\sinh(\om N/2)}
 = \sum_{n=0}^{m-1}\frac{\cosh\big[\om(\tau+nN+mN/2)\big]}{\sinh(\om mN/2)}\,,
\end{equation}
where
\begin{equation}
 T=\frac{1}{a_\tau N},\,\, T_r = \frac{1}{a_\tau N_r},\quad 
\frac{N_r}{N} = m \in \mathbb{N}\,,
\end{equation}
we find
\begin{equation}
G_r(\tau;T,T_r) = \sum_{n=0}^{m-1}G(\tau+nN,T_r)\,.
\label{eq:Grec-direct}
\end{equation}
In this study we will use $T_r=44\,\mrm{MeV} (N_\tau=128)$ as the
reference temperature throughout, and will employ
\eqref{eq:Grec-direct} to compute the reconstructed correlator,
padding $G(\tau;T_r)$ with zeros in the middle where necessary.

\section{Results}
\label{results}

\begin{figure}[thb]
\centering
\includegraphics*[width=0.7\textwidth,clip]{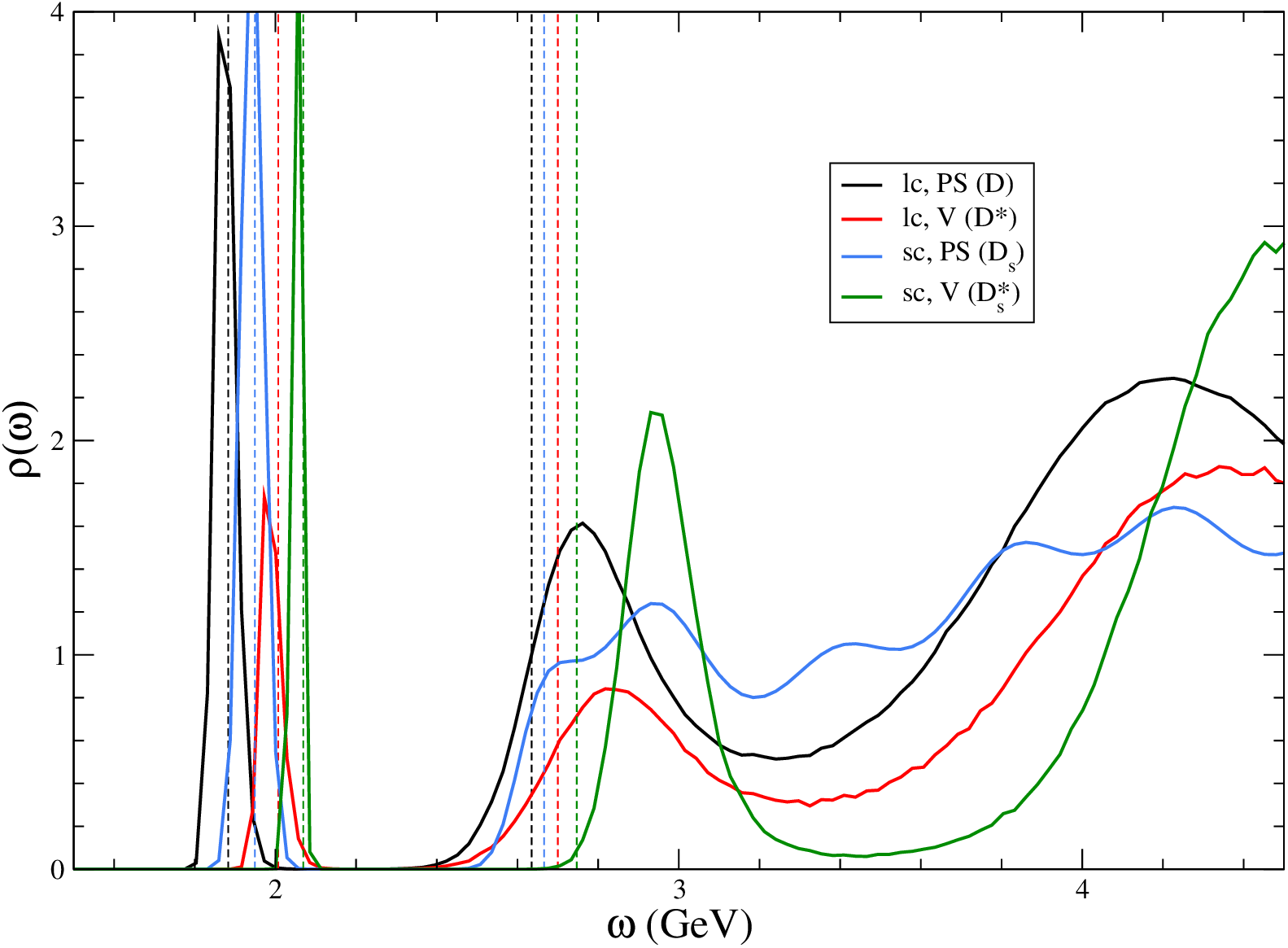}
\caption{Zero temperature spectral function $\rho(\om)$ in all four
  channels.  The vertical lines show the ground states and first
  excited states computed by the HadSpec Collaboration
  \cite{Moir:2013ub}.}
\label{fig:T0}
\end{figure}

We have computed correlators and spectral functions in the
pseudoscalar and vector channels for light--charm ($D, D^*$) and
strange--charm ($D_s, D_s^*$) mesons.  Figure~\ref{fig:T0} shows the
spectral functions at our lowest temperature in all four channels.
We also show the ground and first excited states computed by the
Hadron Spectrum Collaboration~\cite{Moir:2013ub} using a variational
analysis.  Comparing the two, we see that the ground state is very
well reproduced by the MEM, while we are not able to reproduce the
first excited state accurately with our current data.  The wiggles
seen in the $D_s$ spectral function are an artefact of the Fourier
basis.

\begin{figure}[thb]
\centering
\includegraphics*[width=0.9\textwidth,clip]{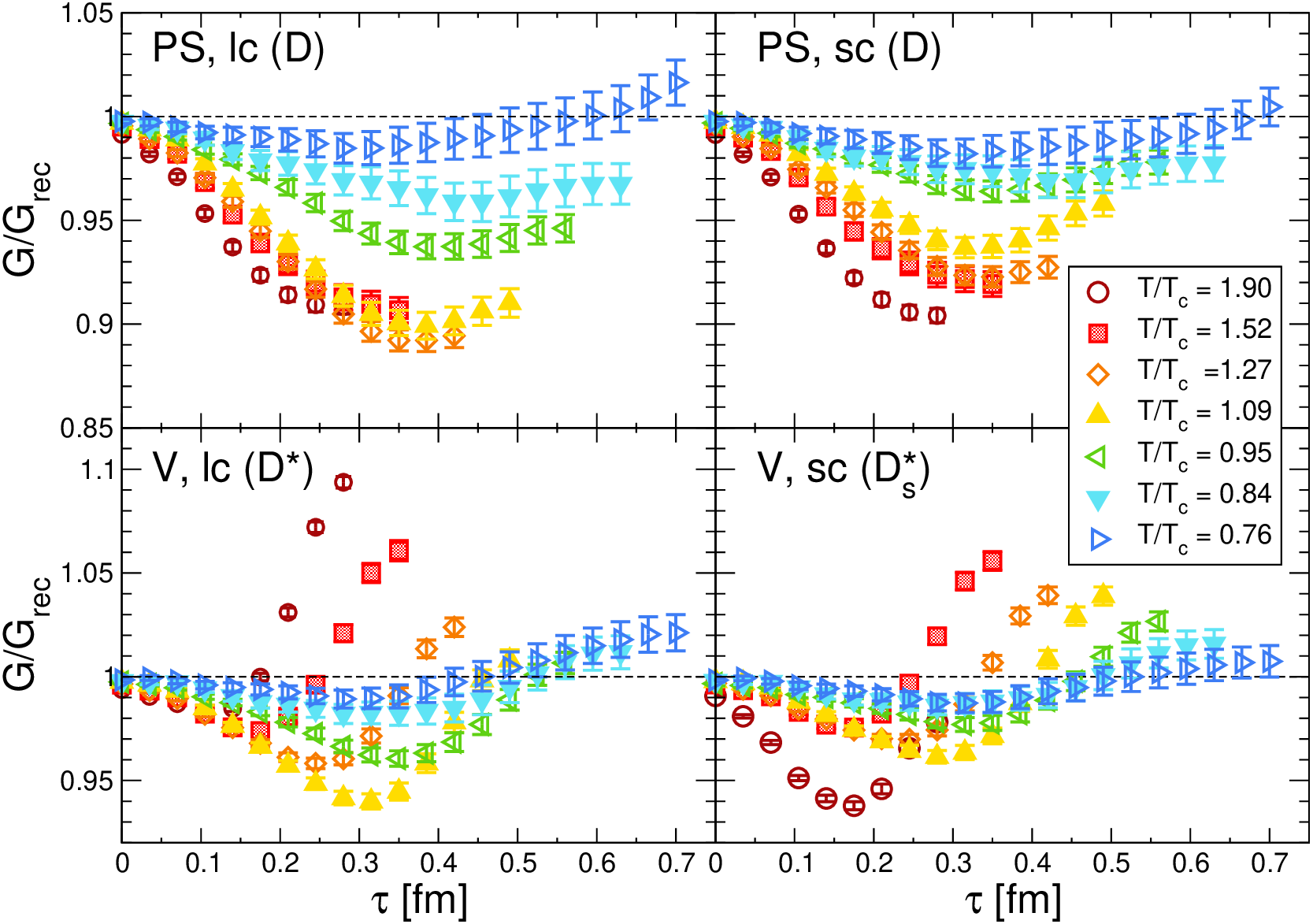}
\caption{Open charm correlators $G(\tau;T)$ divided by the reconstructed
correlator $G_r(\tau;T,T_r)$ computed from eq.~\protect\eqref{eq:Grec-direct}
with $T_r=0.24T_c$.}
\label{fig:Grec}
\end{figure}

In figure~\ref{fig:Grec} we show the correlators $G(\tau)$ divided by
the reconstructed correlators $G_r(\tau)$ at the same temperature, for
all four channels.  This ratio should be 1 if there are no thermal
modifications.  We observe that the correlator at $T=0.76T_c$
is consistent with no modifications in all channels, but there are thermal
modifications already at $T=0.84T_c$, in particular in the $D$ meson
channel.  Above $T_c$, the modifications become significant in all
channels.  We also see that except for the highest temperature
($T=1.9T_c$), the modifications in the strange--charm sector are
smaller than those in the light--charm sector, which may lend support
to the hypothesis that $D_s$ yields may be increased relative to $D$
yields.

\begin{figure}[thb]
\centering
\includegraphics*[width=0.48\textwidth,clip]{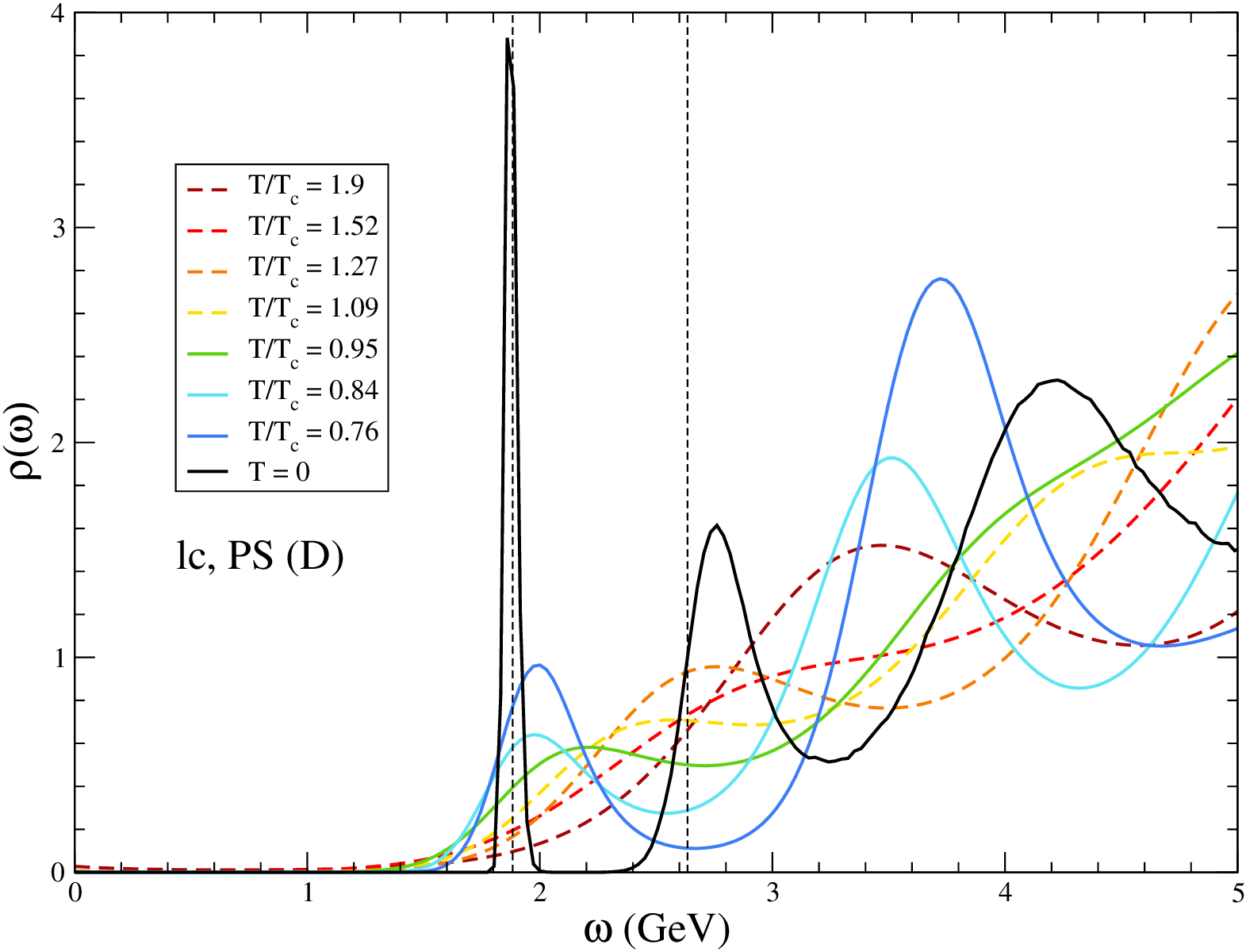}
\includegraphics*[width=0.48\textwidth,clip]{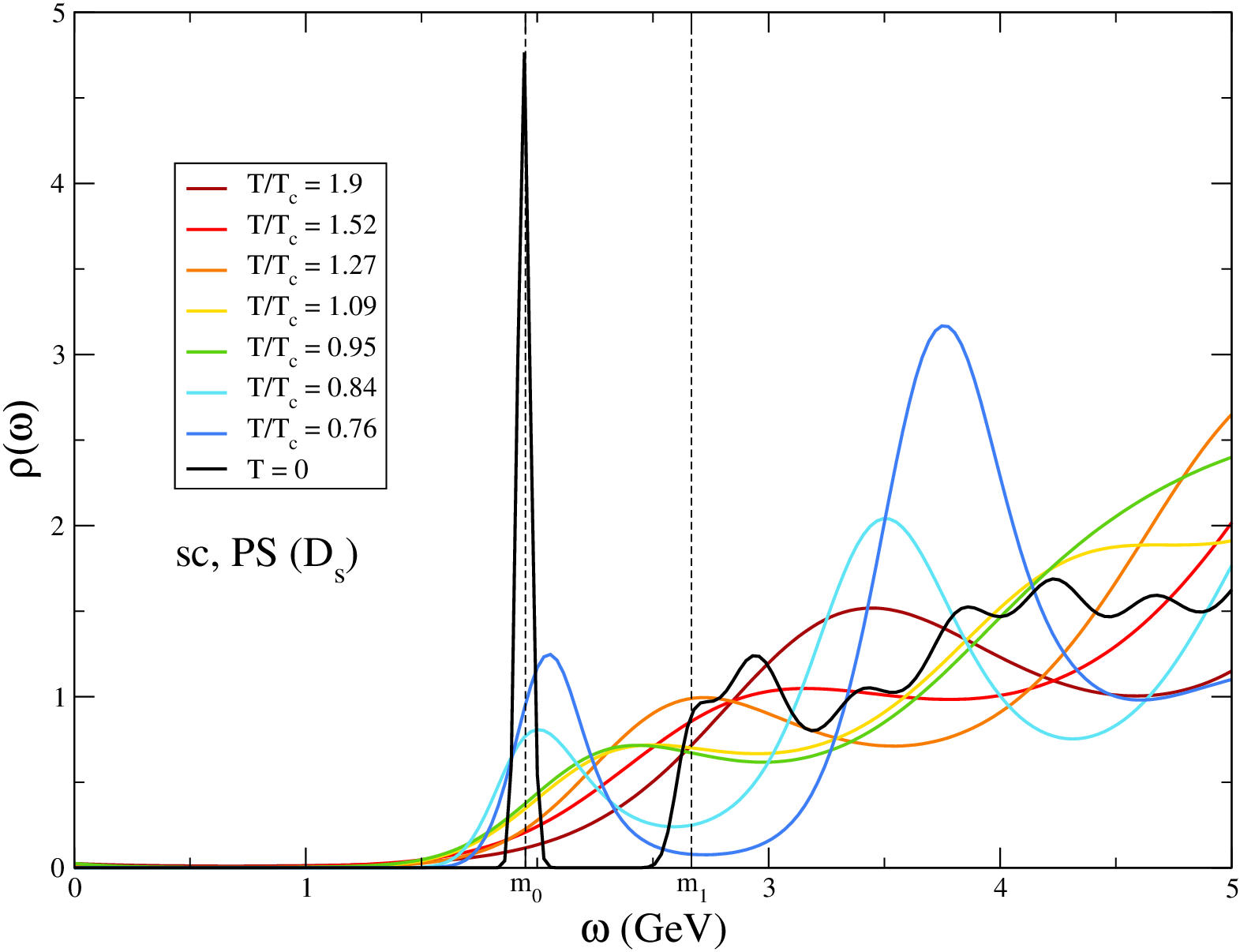}
\caption{Pseudoscalar channel spectral function $\rho(\om)$ for
  light--charm (left) and strange--charm (right) mesons.  The vertical
  lines denote the ground and first excited state masses from
  ref.~\cite{Moir:2013ub}.}
\label{fig:rho-PS}
\end{figure}

\begin{figure}[thb]
\centering
\includegraphics*[width=0.48\textwidth,clip]{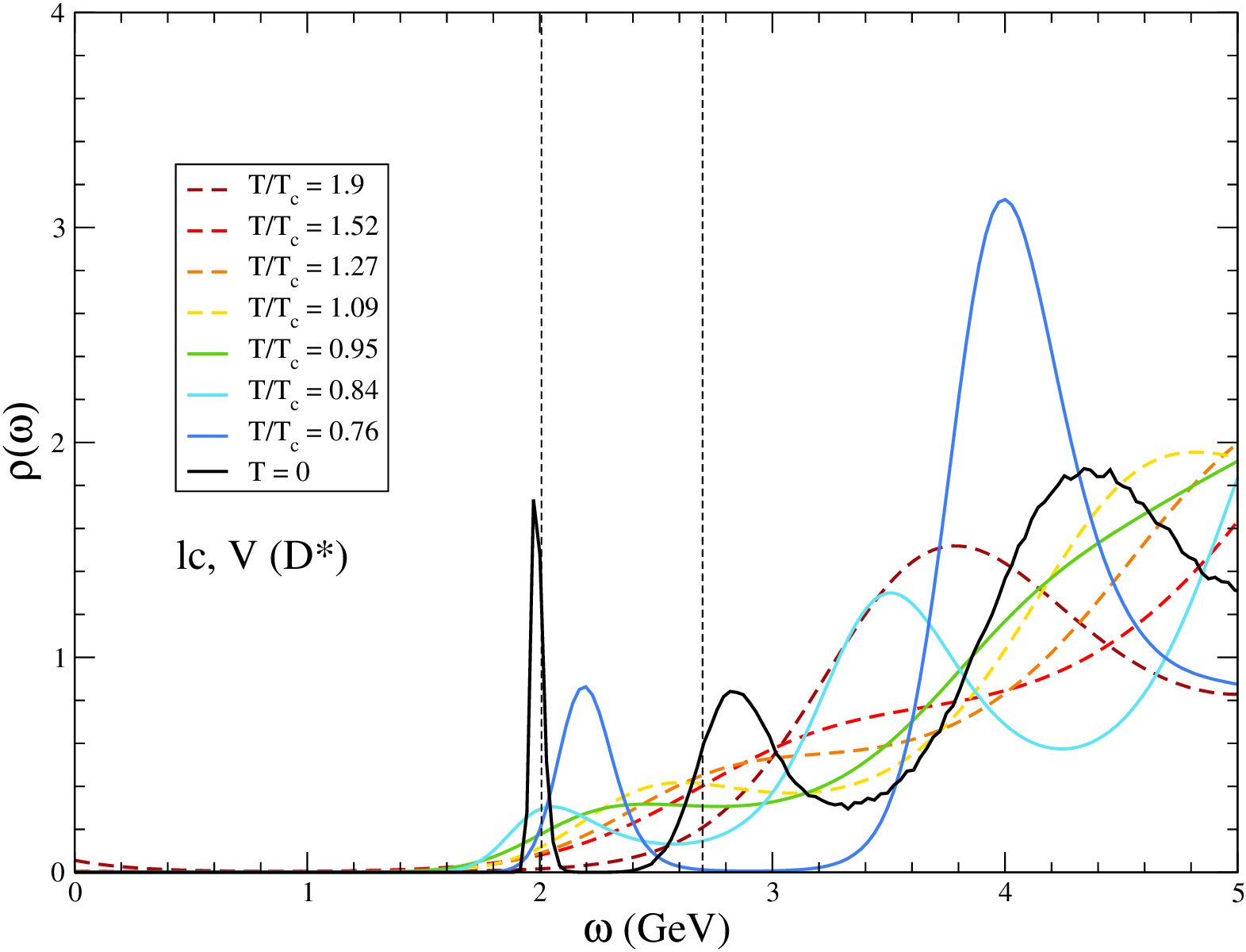}
\includegraphics*[width=0.48\textwidth,clip]{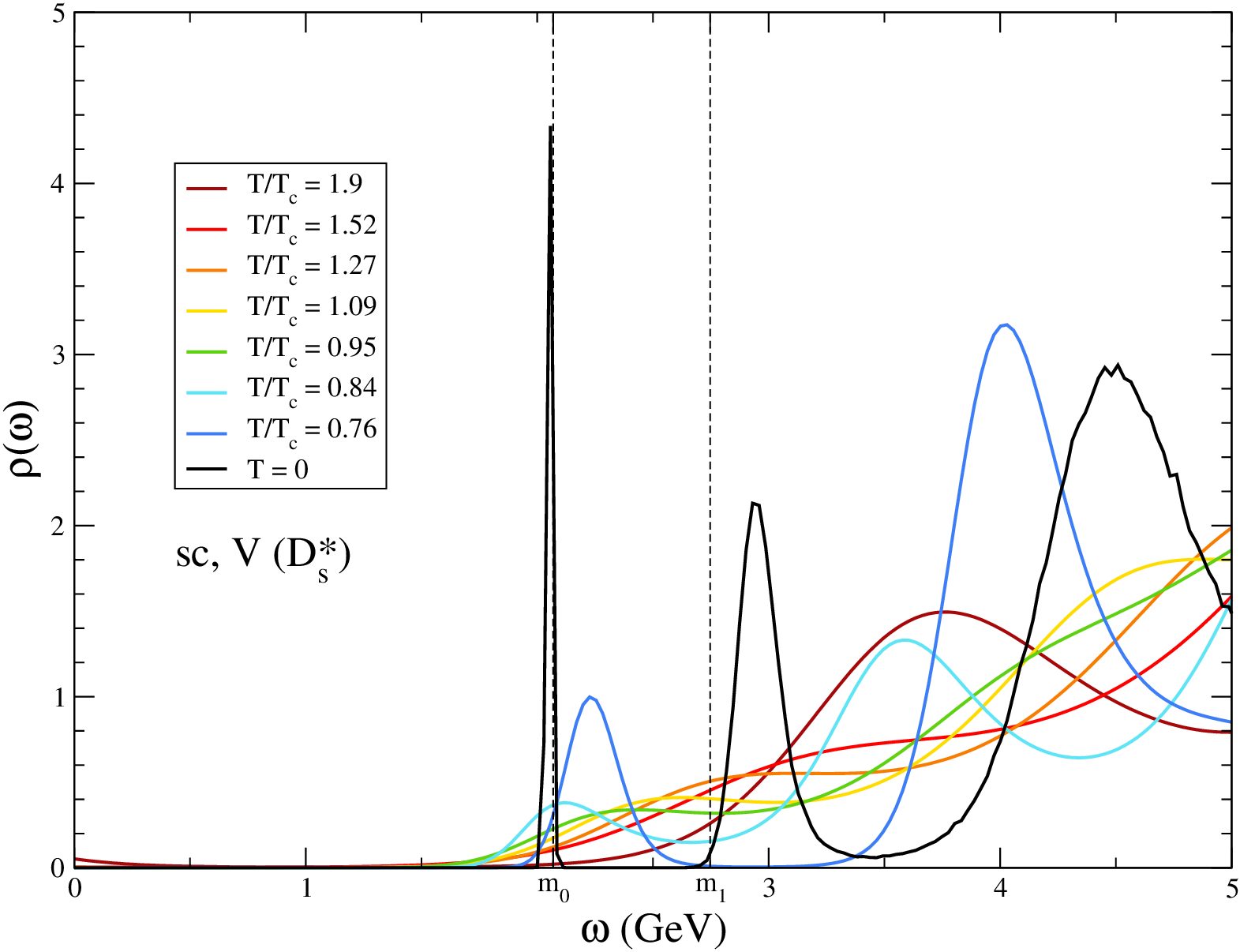}
\caption{Vector channel spectral function $\rho(\om)$ for
  light--charm (left) and strange--charm (right) mesons.  The vertical
  lines denote the ground and first excited state masses from
  ref.~\cite{Moir:2013ub}.}
\label{fig:rho-V}
\end{figure}

We now turn to the spectral functions, which are shown in
figures~\ref{fig:rho-PS} and \ref{fig:rho-V} for the pseudoscalar and
vector channels respectively. We see no sign of any surviving bound
state above $T_c$, suggesting that all open charm states dissociate
near $T_c$.  It may appear that the dissociation occurs already at
$T=0.95T_c$, but it should be noted that the transition to the
quark--gluon plasma
is a broad crossover for our parameters, and that $T=0.95T_c$ lies
within the crossover region~\cite{Aarts:2014nba}. 

Interestingly, there are suggestions, in particular in
the vector channel, of a thermal mass shift below $T_c$, with the
ground state peak position at $T=0.76T_c$ being considerably higher
than at zero temperature.  So far, this feature appears to be robust
with respect to variations in the basis functions and default model,
but work is still in progress to confirm this.

As will be apparent from fig.~\ref{fig:Grec}, the statistical
uncertainties in our correlators are relatively large, at
$\order(1\%)$.  Typically, permille errors or smaller are required to
unambiguously identify spectral features.  The fairly broad features
seen in figs~\ref{fig:rho-PS} and \ref{fig:rho-V} for $T\gtrsim T_c$
may hence be a reflection of our limited precision.  We are currently
working on improving on this by using multiple sources per
configuration for our correlators.

\section{Summary and outlook}
\label{summary}

We have carried out the first lattice calculation of temporal
correlators and spectral functions of open charm mesons, both above
and below the deconfinement crossover.  We find clear evidence of
thermal modifications already below the crossover, which may include a
thermal mass shift.  Above the crossover, we find no evidence of any
surviving bound states.  The thermal modifications are smaller for
$D_s$ and $D_s^*$ mesons than for $D$ and $D^*$ mesons, which may
point to an enhanced yield of $D_s$ relative to $D$ mesons in heavy
ion collisions, in line with theoretical predictions and emerging
experimental evidence~\cite{Adam:2015jda}.

We are currently working on improving our statistics using multiple
sources, which would enable a more precise and reliable extraction of
spectral functions.  We are also carrying out a complete study of MEM
systematics; so far, our main results do not appear to be affected by
these systematics.  We are also planning to cross-check the MEM
results using alternative methods such as the BR
method~\cite{Burnier:2013nla}.

\section*{Acknowledgments}
We thank Alexander Rothkopf for providing access to his MEM code and
for numerous invaluable discussions.  We also thank Gert Aarts and
Chris Allton for fruitful discussions.  We acknowledge the use of
computational resources provided by ICHEC and the STFC funded DiRAC
facility. 

\bibliography{hot,lattice,jis}

\end{document}